\documentclass[aps,twocolumn,prb,showpacs]{revtex4}

\usepackage[dvips]{color}
\usepackage{graphicx}
\usepackage{epsfig}
\usepackage{dcolumn}
\usepackage{color}
\usepackage{amssymb}
\usepackage[english]{babel}
\usepackage{mathrsfs}
\usepackage{bm}
\usepackage{amsmath}
\usepackage{lscape}

\begin{document}

\title[Optimal antibunching ...]
{ Optimal antibunching in passive photonic devices  \\
based on coupled nonlinear resonators
} 

\author{ Sara Ferretti$^{1}$, Vincenzo Savona$^2$, and Dario Gerace$^{1}$ }
\email{dario.gerace@unipv.it} 
\address{$^1$ Dipartimento di Fisica, Universit\`a degli Studi di Pavia, I-27100 Pavia (Italy) \\
  $^2$ Institute of Theoretical Physics, Ecole Polytechnique F\'{e}d\'{e}rale de Lausanne (EPFL), 
                           CH-1015 Lausanne (Switzerland)}


\begin{abstract}
We propose the use of weakly nonlinear passive materials for prospective applications in integrated 
quantum photonics. It is shown that strong enhancement of native optical nonlinearities 
by electromagnetic field confinement in photonic crystal resonators can lead to 
single-photon generation only exploiting the quantum interference of two coupled modes 
and the effect of  photon blockade under resonant coherent driving. 
For realistic system parameters in state of the art microcavities, the efficiency of such single-photon 
source is theoretically characterized by means of the second-order correlation function at zero time 
delay as the main figure of merit, where major sources of loss and decoherence are taken into account 
within a standard master equation treatment. 
These results could stimulate the realization of  integrated quantum photonic devices 
based on non-resonant material media, fully integrable with current semiconductor technology and
matching the relevant telecom band operational wavelengths, as an alternative to single-photon 
nonlinear devices based on cavity-QED with artificial atoms or single atomic-like emitters.
\end{abstract}

\maketitle

\section{Introduction}

Emerging quantum photonic technologies mostly rely on the ability to 
generate, manipulate, and detect quantum states of light, with the challenging 
goal of bringing quantum information and communication devices to large
scale applications (see \cite{qp_review,focus_iqo} for recent overviews).
Single-photon sources (SPS) and linear optical operations are necessary 
ingredients for prospective developments in quantum information processing 
\cite{klm_protocols,mete2004}, and remarkable progress toward a fully integrated 
and CMOS-compatible technology has been achieved in the last few years 
\cite{politi08sci}. However, the ability to generate pure single-photon Fock states 
on demand in a fully integrated quantum photonic platform is still lacking. 
Commonly employed SPS in quantum optical experiments rely on attenuated 
laser beams and parametric down-conversion. Typically, a coherent source of radiation is sent
through a nonlinear medium possessing an intrinsic $\chi^{(2)}$ 
nonlinear susceptibility \cite{hong86prl}, which is able to produce entangled photon
pairs and hence heralded single-photon states upon projective measurement of
one of the two channels \cite{Fasel2004}. Even if such sources inevitably suffer from low 
efficiencies \cite{focus_sps},
this is currently the preferred way to inject single-photon wave packets 
as input for integrated quantum circuits through external fibre coupling \cite{politi08sci},
although an integrated source of heralded single photons has been recently realized \cite{kartik2012}. 

Generation of pure single-photon states requires a quantum nonlinear source, i.e. ideally a two-level emitter 
that is prepared in its excited state and produces exactly a single quantum of light per 
pulse when relaxing to  its ground state. One of the measures of SPS efficiency is determined 
by the degree of antibunching in its second-order correlation function at zero-time delay, 
$g^{(2)}(0)<g^{(2)}(\tau)$, whereby an ideal SPS satisfies $g^{(2)}(0) \to 0$ \cite{loudon_book}.
Tremendous progress in solid-state SPS has been made since the development
of artificial quantum emitters, such as semiconductor quantum dots or atomic-like
defects in solids (for a review, see Ref.~\cite{focus_sps} and references therein). 
In practice, cavity quantum electrodynamics (CQED) is the most straightforward way 
of obtaining single-photon nonlinear behavior and hence a SPS on demand \cite{kuhn_review},
allowing to efficiently access the underlying anharmonicity introduced by a single 
atomic-like emitter through a high-finesse resonator \cite{tian92,imamoglu99}. 
In solid-state CQED, the most promising results on single-photon emission 
have been originally achieved  with III-V materials, by using non-resonantly excited 
semiconductor quantum dots coupled to photonic microcavity modes, 
either in the Purcell regime \cite{Michler2000,Pelton2002,Santori2002},  in strong 
light-matter coupling \cite{kevin07nat,press2007prl}, or even through a direct coupling
to a one-dimensional mode such as a nanowire \cite{claudon2010} or a photon crystal 
waveguide \cite{hughes2007prl,shields2011apl,finley2012prx}.
Single-photon nonlinear behavior in the strong light-matter coupling regime has been 
shown also under resonant excitation \cite{faraon08nphys,volz2012nphot}, i.e. exploiting 
the photon blockade effect \cite{imamoglu97}. 
The latter occurs in systems where two photons inside a 
resonant cavity produce an anharmonic shift of its response frequency that is greater than 
the line-broadening induced by losses and decoherence.
Under such conditions, the transmission of a coherent light field is inhibited 
by the presence of a single-photon within the cavity, until such single-photon state is
released: the system converts a low-power coherent state into single-photon 
streams at the same resonant frequency, thus behaving as a single-photon source. 
The efficiency of such a source can be assessed by measuring the antibunched $g^{(2)}(0)$ 
in the transmitting channel, as experimentally shown for the first time with single caesium 
atoms strongly coupled to an optical Fabry-P\'erot cavity \cite{birnbaum05nat}. 

A similar device can be realized when using a Kerr-type nonlinear medium, which 
gives rise to an equivalent response in terms of  the degree of photon antibunching of the 
transmitted field \cite{rebic04pra}. 
Theoretical proposals have been presented in the last few years to achieve 
single-photon blockade in solid-state systems with \textit{resonant} Kerr-type nonlinearities, 
such as fully confined polariton boxes below the diffraction limit
\cite{ciuti06prb,gerace_josephson,carusot2010epl,ferretti2010,savona10prl,bamba}.
In the latter, the Coulomb interaction of electron-hole pairs in a semiconductor or 
insulating resonant medium is responsible for a Kerr-type nonlinear coefficient, 
potentially able to produce photon blockade. 
Owing to experimental difficulties in reaching diffraction-limited polariton confinement with sufficiently
large cavity lifetime, no evidence of quantum nonlinear behavior from these systems has been reported, yet.
Besides this, applications of such devices as SPS in integrated quantum photonics technologies 
might be limited in the long term by lack of suitable active media in the telecom-band, necessary to be 
interfaced with long-distance communication networks, and the commonly cryogenic 
working temperatures \cite{focus_sps}.

The possibility to engineer a SPS based on single-photon blockade from 
\textit{non resonant} Kerr-type nonlinearities in passive devices (such as
nano-structured silicon or GaAs platforms in the near infrared) could represent 
a breakthrough from quantum photonics perspectives.  
First, it would allow large flexibility on the operational wavelengths, to be 
engineered in the telecom band by the photonic confinement in the materials 
transparency window. 
Second, it would potentially allow room temperature operations,
and possibly the elevated temperatures required for telecoms devices.
Finally, it would only require a coherent pump source such as a standard
fiber-coupled telecom laser, thus providing a compact, stable, and relatively 
low cost device.
It is a common perception that ordinary nonlinear media are intrinsically 
unable to display significant single-photon nonlinear behavior and a very large 
number of photons is necessary to produce appreciable effects, owing to the 
small value of the $\chi^{(3)}$ elements in ordinary bulk media \cite{boyd_book}.
However, interesting perspectives might come from a proper nano-structuring of 
passive nonlinear materials, which could lead to strongly enhanced effective nonlinearities 
down to the single-photon level \cite{ferretti2012}, especially using photonic crystal type
confinement allowing to reach ultra-high quality factors \cite{high_Q} and ultra-small confinement
volumes \cite{arakawa2011} in a purely dielectric structure. Even if potentially not unrealistic, such 
goal is still quite challenging with single resonators. 

Here we report on an alternative strategy to realize an integrated
SPS based on non-resonant Kerr-type nonlinearity, which exploits the
effects of quantum interference in coupled photonic resonators
\cite{savona10prl,bamba}. 
These effects have been shown to arise even for weak
nonlinear response, namely for an anharmonic shift of the two-photon energy
levels much smaller than the cavity mode linewidth. Though very
small, this anharmonic shift is enough to introduce a phase shift
between different excitation pathways to the two-photon state, which
for well chosen values of the system parameters suppresses its
occupation as a result of destructive interference, leading to antibunching
of the output light.  
We notice that an analogous mechanism had already been characterized in 
the simplest CQED system -- a two-level atom coupled to a single mode resonator -- 
in a seminal work by H. Carmichael \cite{carmichael85}. 
For a given value of the anharmonic shift, this scheme is able to relax by more than
three orders of magnitude the cavity loss rate over 
conventional photon blockade devices, making the mechanism perfectly viable, 
for example, with commonly achievable defect cavities in photonic crystal structures.
For such structures, we further show that the photon antibunching is robust to 
typically expected rates of pure dephasing. A generalization of this scheme to 
multi-cavity arrays might give rise to interesting perspectives for the generation
of continuous variable entangled states on chip \cite{liew2012}.
{ In the present work  we concentrate our analysis on the optimization of  
$g^{(2)}(0)$ as a figure of merit for SPS, leaving the specific characterization of 
the device output in terms of single-photon wave packets for future reports.
With this, we also expect this proposal to stimulate new research efforts in
silicon quantum photonics and CMOS-compatible quantum information technology.}

The paper is organized as follows. In section \ref{sec:model} we present the basic model, based on 
a master equation for the system hamiltonian and dissipative terms including the effects of 
losses and dephasing. In section \ref{sec:results}, we give an overview of the steady state results,
quantifying the optimized antibunching as a function of the system parameters, and test its 
robustness with respect to pure dephasing mechanisms. 
Finally, we draw the conclusions of this work in section \ref{sec:conclu}.

\section{Model of driven/dissipative coupled nonlinear resonators}
\label{sec:model}

\begin{figure}[t]
\begin{center}
\vspace{-0.5 cm} 
\includegraphics[width=0.5\textwidth]{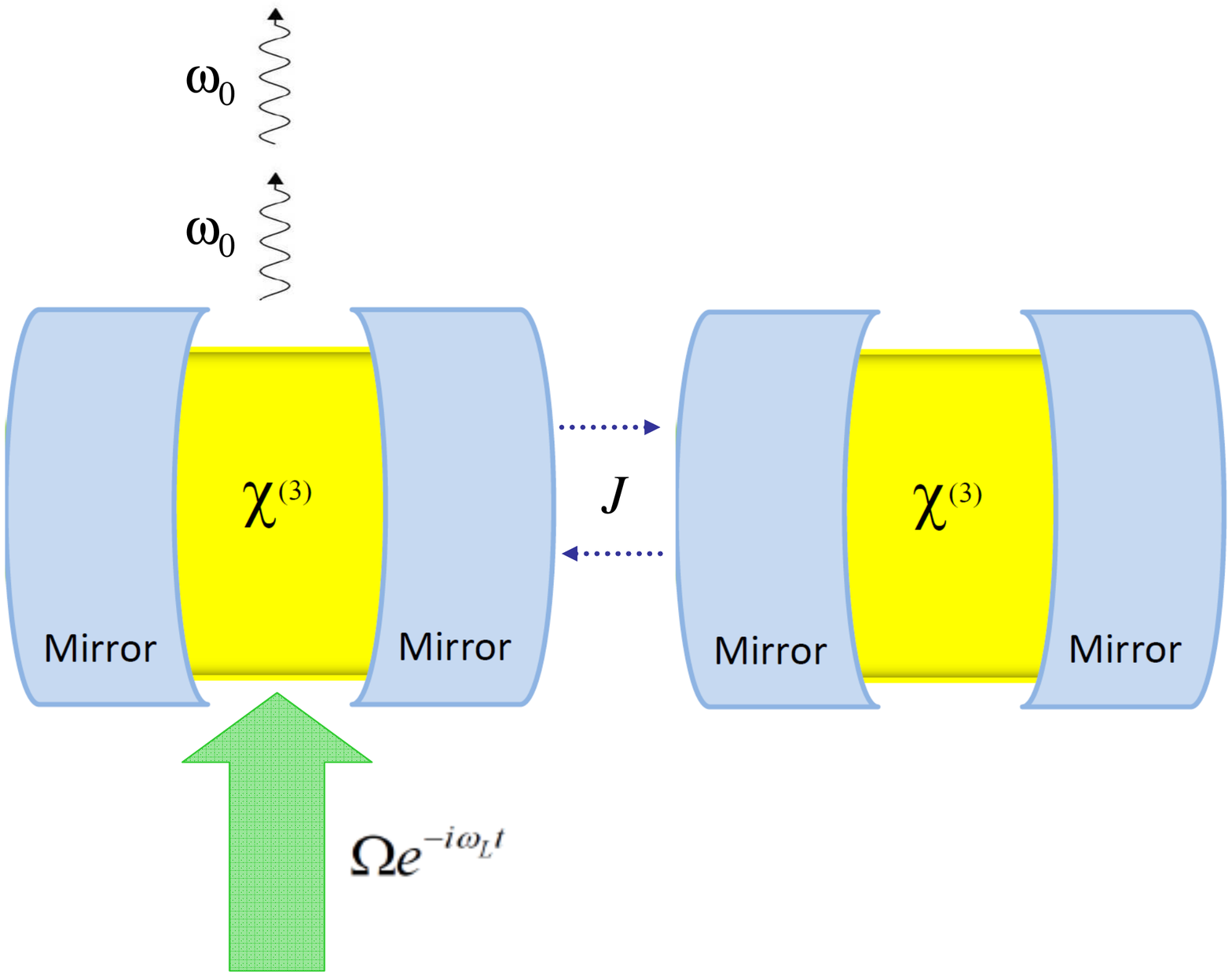}
\vspace{-0.8 cm} 
 \caption{Scheme of two coupled resonators made of a Kerr-type 
  nonlinear material. We highlight that just one of the cavities is resonantly driven by a coherent 
  light source, and output light is collected from the same according to the scheme proposed in \cite{savona10prl}.} 
  \label{sistema}
\end{center}
\end{figure}

The formalism introduced in the previous work for a single 
resonator \cite{ferretti2012}  can be straightforwardly generalized to the case of two tunnel-coupled 
coupled resonators made of a  Kerr-type nonlinear material, or \textit{photonic molecule}, schematically 
represented in figure~\ref{sistema}. 
Each resonator is assumed to be single-mode, where $\hat{a}_i$ ($\hat{a}_i^{\dagger}$) 
define the destruction (creation) operator of a single photon ($i=1,2$ labels the two resonators). 
As in \cite{savona10prl,bamba}, we consider the nonlinear second-quantized hamiltonian  
\begin{eqnarray}
\hat{H}=\sum_{i=1,2}[\hbar\omega_{i} \hat{a}_{i}^{\dag}\hat{a}_{i}+
U_{\mathrm{nl}}\hat{a}_{i}^{\dag}\hat{a}_{i}^{\dag}\hat{a}_{i}\hat{a}_{i}] + \nonumber \\
+ J(\hat{a}_{1}^{\dag}\hat{a}_{2}+\hat a_{2}^{\dag}\hat{a}_{1})+F
e^{-i\omega_L t} \hat{a}_{1}^{\dagger}+F^{\ast} e^{i\omega_L
t}\hat{a}_{1} \, ,
\end{eqnarray}
where the linear part describes two harmonic oscillators, neglecting 
the zero point energy, $J/ \hbar$ is the tunnel coupling rate between the 
two resonators \cite{nota1}, 
$F/ \hbar$ is the coherent pump rate at the continuous wave (cw) laser frequency $\omega_L$.
Following the scheme introduced in \cite{savona10prl}, only one of the two resonators 
is driven, and light will be assumed to be collected only from the same.
For the case of passive nonlinear resonator, the nonlinear energy shift for each mode can 
be approximated as \cite{drummond80,ferretti2012} 
\begin{equation}\label{shift_kerr}
U_{\mathrm{nl}}\simeq \frac{3(\hbar\omega_0)^2 } {4\varepsilon_0}
\frac{\overline{\chi}^{(3)}}{\overline{\varepsilon}_r^2}
\int  |\vec{\alpha}_i(\mathbf{r})|^4 \mathrm{d}\mathbf{r} 
=  \frac{3(\hbar\omega_0)^2 } {4\varepsilon_0 V_{\mathrm{eff}}}
\frac{\overline{\chi}^{(3)}}{\overline{\varepsilon}_r^2} \, ,
\end{equation}
where the effective mode volume is defined as 
$V^{-1}_{i,\mathrm{eff}}=V^{-1}_{\mathrm{eff}}=\int  |\vec{\alpha}_i(\mathbf{r})|^4 \mathrm{d}\mathbf{r}$,
from the normalized field profile in each cavity, $\vec{\alpha}_i$, for 
the specific nanocavity implementation (see \cite{vahala_review} for a review).
To this purpose, we will assume constant values for the average 
real part of the nonlinear susceptibility and relative dielectric permittivity, $\overline{\chi}^{(3)}$ 
and $\overline{\varepsilon}_r$ respectively, to have realistic order of magnitude estimates on
materials of current technological interest \cite{nota2}.
The hamiltonian can be expressed in a rotating frame with respect to the pump
frequency, $\Delta\omega_{i}=\omega_{i}-\omega_{L}$,
\begin{eqnarray}\label{eq:model}
\hat{H}=\sum_{i=1,2}[\hbar\Delta\omega_{i} \hat{a}_{i}^{\dag}\hat{a}_{i}+
U_{\mathrm{nl}}\hat{a}_{i}^{\dag}\hat{a}_{i}^{\dag}\hat{a}_{i}\hat{a}_{i}]
&+& J(\hat{a}_{1}^{\dag}\hat{a}_{2}+\hat a_{2}^{\dag}\hat{a}_{1}) \nonumber \\
&+&F\hat{a}_{1}^{\dagger}+F^{\ast}\hat{a}_{1} \, .
\end{eqnarray}

The driven/dissipative character of this model is taken into account by a master equation treatment
for the system density matrix in Markov approximation
\begin{equation}\label{rho2res}
\dot{\rho}=\frac{1}{i \hbar}[\rho,\hat{H}] + \mathcal{L}^{(1,2)} + \mathcal{L}^{deph}\, ,
\end{equation}
where
\begin{equation}
\mathcal{L}^{(1,2)}= \sum_{i=1,2} \frac{\gamma_{i}}{2}  [2\hat{a}_{i}\rho
\hat{a}_{i}^{\dag} -
\hat{a}_{i}^{\dag}\hat{a}_{i}\rho-\rho\hat{a}_{i}^{\dag}\hat{a}_{i}] \, ,
\end{equation}
is the Liouvillian operator in the usual Lindblad form for the two resonators modes,
and 
\begin{equation}
\mathcal{L}^{deph}= 
              \sum_{i=1,2} \frac{\gamma_i^{\ast}}{2}  [2\hat{a}_{i}^{\dag}\hat{a}_{i}\rho\hat{a}_{i}^{\dag}\hat{a}_{i}
            - \hat{a}_{i}^{\dag}\hat{a}_{i} \hat{a}_{i}^{\dag}\hat{a}_{i} \rho
            - \rho\hat{a}_{i}^{\dag}\hat{a}_{i} \hat{a}_{i}^{\dag}\hat{a}_{i} ] \, ,
\label{eq:deph12}
\end{equation}
models the pure dephasing for the two resonators modes.

In this work, we generalize the model to account for possible unbalanced losses between the
two normal modes of the system. 
In the following, we will only consider the case $\Delta\omega_{1}=\Delta\omega_{2}$
(with straightforward generalization), under which conditions one has the two symmetric and antisymmetric
combinations, respectively given by
\begin{equation}
\hat{a}_{+}=\frac{1}{\sqrt{2}}(\hat{a}_{1}+\hat{a}_{2})\, ; \, \, \, \,
\hat{a}_{-}=\frac{1}{\sqrt{2}}(\hat{a}_{1}-\hat{a}_{2})\, .
\end{equation}
In this case, the losses can be described through the corresponding Liouvillian for the
two coupled modes in the Lindblad form
\begin{eqnarray}
\mathcal{L}^{(\pm)}= &&\frac{\gamma_{+}}{2}[2\hat{a}_{+}\rho
\hat{a}_{+}^{\dag} -
\hat{a}_{+}^{\dag}\hat{a}_{+}\rho-\rho\hat{a}_{+}^{\dag}\hat{a}_{+}]  \nonumber \\
&&   + \frac{\gamma_{-}}{2}[2\hat{a}_{-}\rho \hat{a}_{-}^{\dag} -
\hat{a}_{-}^{\dag}\hat{a}_{-}\rho-\rho\hat{a}_{-}^{\dag}\hat{a}_{-}]\, \, .
\end{eqnarray}
One can easily show that, when $\gamma_{1}=\gamma_{2}=\gamma$, 
and hence $\gamma_{\pm}=\gamma$, we get 
$\mathcal{L}^{(\pm)}=\mathcal{L}^{(1,2)}$, as expected.
For the latter model, pure dephasing is taken into account by the Lindblad term
as in Eq.~(\ref{eq:deph12}), but for the operators $\hat{a}_{+}$ and $\hat{a}_{-}$.

\begin{figure}[t]
\begin{center}
\vspace{-0.4 cm} 
\includegraphics[width=0.5\textwidth]{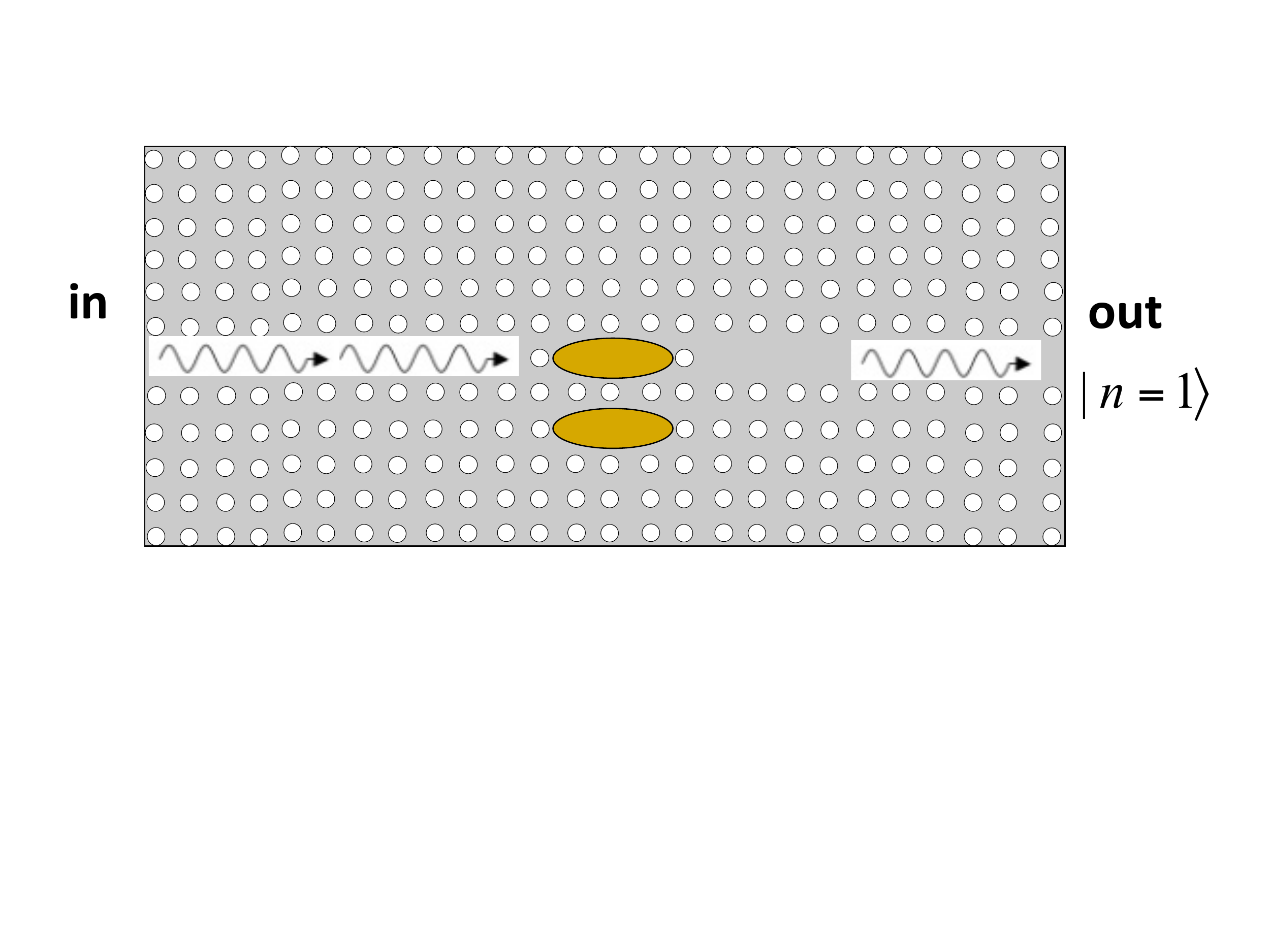}
\vspace{-2.5 cm} 
 \caption{
Sketch of a possible realization of a SPS based on the scheme shown in figure \ref{sistema}, with
a photonic crystal integrated circuit fabricated with a nonlinear material as the underlying platform.}  
  \label{schema}
\end{center}
\end{figure}

\section{Towards passive single-photon sources}
\label{sec:results}

We hereby analyze the device represented in the scheme of figure~\ref{sistema} as a potential 
SPS:  a photonic molecule is driven by a low power input laser 
impinging only on cavity 1, and output light is collected from the same cavity (see next section
for a possible realization). The device could realize an efficient SPS if the light
emitted from cavity 1 is strongly antibunched \cite{savona10prl}. In the following, 
we assume the normalized degree of second order coherence from the first resonator, 
$g^{(2)}(0) = \langle\hat{a}_1^{\dag2}\hat{a}_1^{2}\rangle/ \langle\hat{a}_1^{\dag}\hat{a}_1\rangle^2$,
as the figure of  merit quantifying the device efficiency as SPS.

\subsection{Model parameters }

We consider passive materials where the non-resonant nonlinearity is
determined by the bulk contribution to the third-order material susceptibility,
where the real part of the $\chi^{(3)}$ elements is responsible for the nonlinear frequency 
shift, while the imaginary part gives rise to an additional contribution to losses,
such as two-photon absorption (TPA). As shown in \cite{ferretti2012}, TPA is 
essentially negligible for the low power intensities that would be required for
single-photon nonlinear operation in such devices. For this reason, we will
neglect its effect here. Typical semiconductor or insulator materials employed
in the optoelectronic industry possess a real part of third-order susceptibility 
of the order of $10^{-19} - 10^{-18}$ in Si units (m$^2/$V$^2$) \cite{boyd_book,ferretti2012}. 
We focus here on photonic crystal type confinement, where remarkable figures of merit 
(such as, e.g., ultra-high quality factors) have
been already shown both with silicon (for a review, see \cite{notomi_review}) and 
GaAs materials \cite{derossi2008}. Moreover, strong enhancement of second and
third order nonlinearities has been also reported \cite{derossi2008,galli2010,graphene2012}.
In such nanostructures,  a realistic cavity mode confinement can be on the order 
of $V_{\mathrm{eff}}=(\lambda / n_r)^3\simeq 0.1$ $\mu$m$^3$ (for near infrared 
operation, i.e. $\lambda\sim 1$ $\mu$m  and $n_r=3$ for typical semiconductors in this 
wavelength range), 
from equation~(\ref{shift_kerr}) we have an estimate for the single-photon nonlinearity
on the order of $U_{\mathrm{nl}}\simeq 10^{-3}$ $\mu$eV. This is certainly a very
small value, as compared to typical resonant nonlinearities induced by QD or QW
transitions \cite{ciuti06prb}.
However, we aim at exploiting the photon blockade mechanism in weakly nonlinear
tunnel-coupled devices, as introduced in \cite{savona10prl}. To this end, we assume a very 
specific realization of such proposal involving coupled photonic crystal cavities and
access waveguides integrated on the same photonic chip, as schematically represented
in figure \ref{schema}. 
In this case, typical tunnel coupled micro-cavities display modes with spectral splitting on the order
of $\Delta E_{\pm}=2J\sim 1$ to few meV \cite{atlasov08oe,caselli2012}.
Finally, the quality factor of cavity modes in the near infrared is typically 
on the order of $Q\sim 10^4$ to $10^5$, and we can safely assume $Q\sim 30000$ as a
realistic value, which gives a typical cavity mode linewidth $\hbar\gamma \simeq 0.025$ meV for a 
resonance around $\hbar\omega_1\simeq 0.8$ eV (the typical telecom wavelength
$\lambda=1.55$ $\mu$m). 
Such realistic values for silicon or GaAs based photonic crystal chips at 
operational wavelengths in the most interesting telecom band allow to
assume normalized parameters in the model hamiltonian (\ref{eq:model}) 
that are consistent with state-of-the art semiconductor-based photonic 
technology: $U_{\mathrm{nl}} \sim 5\times10^{-5}$,
$J\sim 50$ (in units of $\hbar\gamma$).

\begin{figure}[t]
\begin{center}
\includegraphics[width=0.5\textwidth]{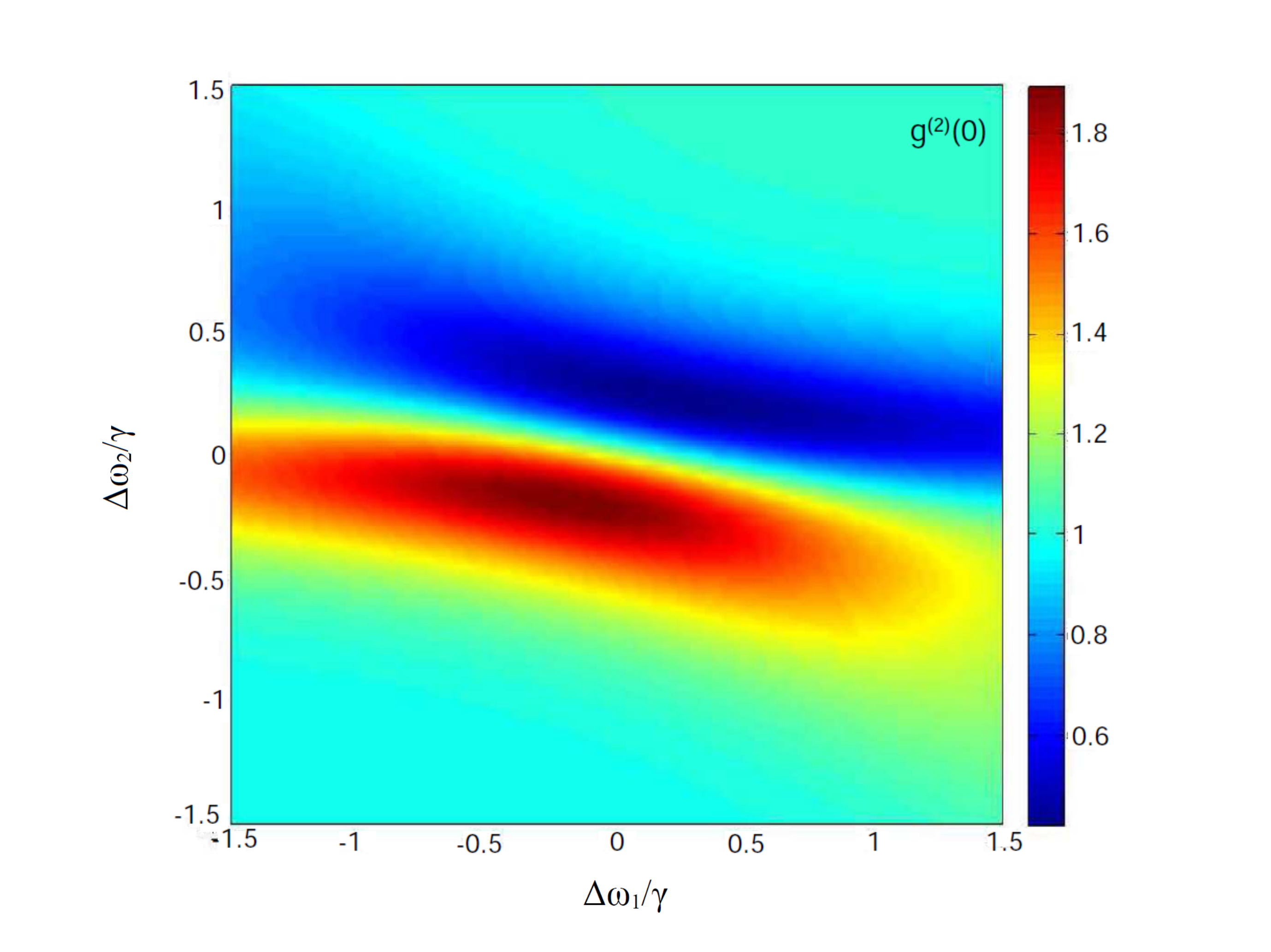}
\vspace{-0.5cm} 
\caption{Numerical solution for the zero-time-delay second-order correlation 
represented in color scale plot as a function of  $\Delta\omega_{1}\neq\Delta\omega_{2}$, with 
parameters (units of $\hbar \gamma$): 
$\hbar \gamma=25$ $\mu$eV, $U_{\mathrm{nl}}=4\times10^{-5} $ 
and $J=56$, $F=10$. } 
\label{fig:omega12}
\end{center}
\end{figure}

\subsection{Optimal antibunching}

Given the basic numbers, we now explore the efficiency of such SPS spanning the 
parameters space for the corresponding master equation, which is solved 
in its steady state by numerically determining a quantum average on $\rho_{ss}$, 
which is the density matrix corresponding to the eigenvalue $\lambda_{ss}=0$ in the linear 
eigenvalue problem  $\cal{L}\rho = \lambda \rho$. 

First, in order to numerically infer the optimal resonance-pump detunings, 
we show in figure~\ref{fig:omega12} the calculated $g^{(2)}(0)$  as a function of $\Delta\omega_{1}$ 
and $\Delta\omega_{2}$, respectively. 
We truncate the Hilbert space to consider a maximum number of photons $N_{\mathrm{max}}=12$ 
in the basis set of Fock states, which we checked to be sufficient for convergence.
Realistic model parameters are assumed, as described in the previous section and 
reported in the figure caption. Again, we point out that all the energies are intended in units of 
$\hbar\gamma = 25$ $\mu$eV, as explained above.
The minimum value of $g^{(2)}(0)$ identifies the optimal operation detuning of the 
laser frequency from the bare resonators, which is  
$\Delta\omega_{1}=\Delta\omega_{2} \simeq 0.2$, for which $g^{(2)}(0)\simeq 0.42$. 

\begin{figure}[t]
\begin{center}
\includegraphics[width=0.5\textwidth]{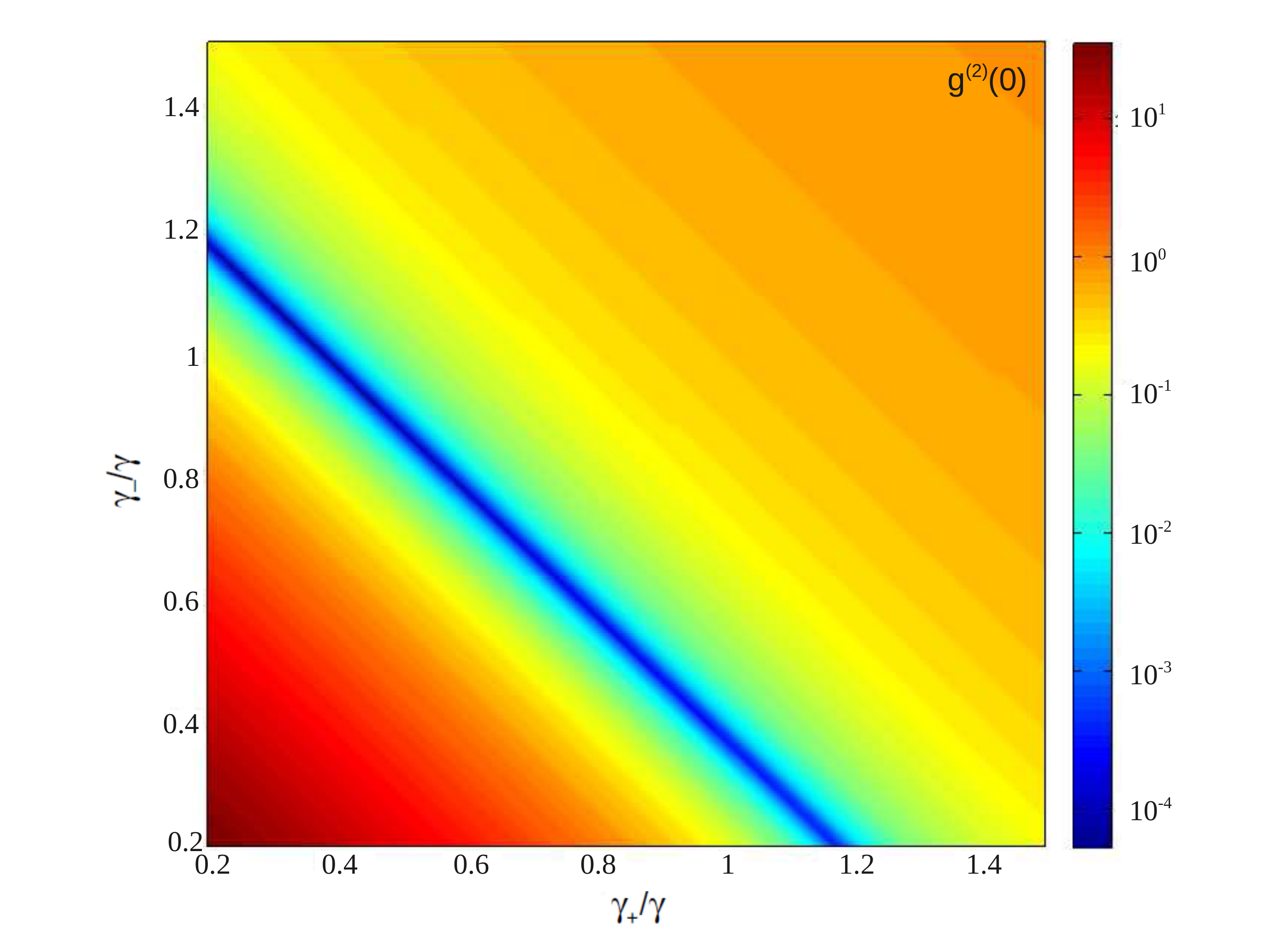}
\caption{
Color scale plot of the numerical solution for the zero time-delay second-order correlation 
as a function of $\gamma_{+}/\gamma$ and $\gamma_{-}/\gamma$, with 
parameters (in units of $\hbar \gamma=25$ $\mu$eV):  
$\Delta\omega_{1}=\Delta\omega_{2}=0.2$; $F=15$;
$U_{\mathrm{nl}}=4\times10^{-5}$, and $J=56$.
} 
 \label{fig:gamma_pm}
\end{center}
\end{figure}

While the previous parameter can be experimentally adjusted with a tunable laser as
input source, there are characteristics that depend on the specific device implementation.
In particular, coupled modes of photonic crystal molecules may display asymmetric
losses according to the mode parity, i.e. $\gamma_{+}\neq \gamma_{-}$, as experimentally shown, e.g., 
in \cite{atlasov08oe}. 
This effect is not surprising, since unbalanced Q-factors of symmetric and antisymmetric modes 
can be traced back to the peculiar loss mechanism in such planar photonic crystal cavities, 
where perturbative coupling to radiative modes depends on the 
mode parity with respect to a defined symmetry plane of the system \cite{andreani06prb}. 
In figure~\ref{fig:gamma_pm}, we consider the case in which the two resonators possess the 
same basic characteristics, i.e. the same frequency detunings 
$\Delta\omega_{1}=\Delta\omega_{2}=0.2$, and the same dissipation rates
for each isolated cavity mode, $\gamma$. We then study the behavior 
of the SPS in terms of the  symmetric and antisymmetric modes losses.
Numerical solutions are reported for the zero-time-delay second-order correlation 
as a function of  $\gamma_{+}/\gamma$ and of $\gamma_{-}/\gamma$.
Quite remarkably, the figure of merit is optimal for unbalanced losses, i.e. 
$\gamma_{+} \neq \gamma_{-}$. 
In fact, breaking the symmetry of the system and for 
appropriate resonators parameters the photonic molecule in 
exam represents an almost ideal source of single-photon Fock states
(down to $g^{(2)}(0) < 10^{-4}$). 
Incidentally, this is exactly the situation that is most naturally realized in a
photonic crystal molecule \cite{atlasov08oe}, although not generally valid for 
any type of tunnel-coupled photonic resonators (such as, e.g., coupled microdisks or 
micropillars). 

{ The numerical results reported in figure~\ref{fig:gamma_pm} can be understood 
by employing a perturbative analytical treatment \cite{ferretti2010,bamba}. 
The pump laser frequency in the present scheme is strongly off-resonance with respect to the 
two eigenfrequencies of the system, corresponding to the $\hat{a}_{+}$ and $\hat{a}_{-}$ modes, 
respectively. Hence, even under the conditions 
$F/ (\hbar\gamma) \gg 1$, a low occupancy of the two modes is expected and the Hilbert space can be truncated 
to the two-photon occupancy in the basis of Fock 
states $\{ |n_1, n_2 \rangle \}$, with $n_1$ and $n_2$ counting the number of photons in the
first and second mode, respectively. 
The modes losses can be treated at the hamiltonian level by considering 
\begin{equation}\label{eq:model_tilde}
\tilde{H}= \hat{H} -i\frac{\hbar\gamma_{+}}{2}\hat{a}^{\dagger}_{+}\hat{a}_{+} -i\frac{\hbar\gamma_{-}}{2}\hat{a}^{\dagger}_{-}\hat{a}_{-}\, ,
\end{equation}
 where $\hat{H}$ is given in Eq.~(\ref{eq:model}). With straightforward substitution, Eq.~(\ref{eq:model_tilde})
 can be written as
 \begin{equation}
 \tilde{H}=\sum_{i=1,2}[\tilde{\Delta}_{i} \hat{a}_{i}^{\dag}\hat{a}_{i}+
U_{\mathrm{nl}}\hat{a}_{i}^{\dag}\hat{a}_{i}^{\dag}\hat{a}_{i}\hat{a}_{i}]
+ \tilde{J}(\hat{a}_{1}^{\dag}\hat{a}_{2}+\hat a_{2}^{\dag}\hat{a}_{1})+F(\hat{a}_{1}^{\dagger}+\hat{a}_{1}) \, .
\end{equation}
 where we have defined $\tilde{\Delta}_i= \hbar[\Delta\omega_{i} - (\gamma_{+}+\gamma_{-})/4]$ and 
 $\tilde{J}=J- \hbar(\gamma_{+}-\gamma_{-})/4$.
 With the ansatz
 \begin{eqnarray}
\vert \psi \rangle=C_{00} \vert 00\rangle &+& C_{10} \vert 10\rangle + C_{01} \vert 01\rangle 
+ C_{20} \vert 20\rangle \nonumber \\
&+& C_{11} \vert 11\rangle + C_{02} \vert 02\rangle + ... \, ,   
\end{eqnarray}
 the steady state solution can be found by solving the coupled equations for the coefficients
 $C_{n_1 n_2}$ from $i\hbar\frac{\partial}{\partial t} \vert \psi \rangle  = \tilde{H} \vert \psi\rangle = 0$
 \begin{eqnarray}
 & \tilde{\Delta} C_{10}+J C_{01} + F C_{00} = 0 \,  \label{eq:lowF1}\\
 & \tilde{\Delta}C_{01}+ J C_{10}  = 0 \,  \\
 & 2 (\tilde{\Delta} + U_{\mathrm{nl}}) C_{20}+\sqrt{2} J C_{11} + \sqrt{2} F C_{10} = 0 \,  \\
 & 2 \tilde{\Delta} C_{11}+\sqrt{2} J (C_{20}+C_{02})+ F C_{01} = 0 \, , \\
 & 2 (\tilde{\Delta} + U_{\mathrm{nl}}) C_{02}+\sqrt{2} J C_{11}  = 0 \, ,\label{eq:lowF5} 
\end{eqnarray}
 where we have approximated $\tilde{J}\simeq J$, since $J\gg \hbar\Delta\omega, U_{\mathrm{nl}}, \hbar\gamma$.
Eqs. (\ref{eq:lowF1})-(\ref{eq:lowF5}) can be simplified by considering the condition 
$C_{00} \simeq 1$ and $C_{00} \gg C_{10}, C_{01} \gg C_{20}, C_{11}, C_{02}$. 
The second-order correlation for light emitted from cavity 1 can be analytically approximated as 
 $g^{(2)}(0)\simeq 2|C_{20}|^2 / |C_{10}|^4$. Hence, minimizing  $|C_{20}|^2 $ 
 with respect to $\Delta\omega_1= \Delta\omega_2 = \Delta\omega$ 
 (keeping fixed $\gamma_{+}+\gamma_{-}$, $J$, $U_{\mathrm{nl}}$), which corresponds to
 minimizing the complex function $|2\tilde{\Delta}^3 + J^2 U_{\mathrm{nl}}|^2$, we get the optimal
 conditions for the antibunching behavior at fixed $\Delta\omega$ as  
 \begin{equation}
 [\gamma_{+}+\gamma_{-} ]_{\mathrm{opt}} \simeq 4\sqrt{3} \Delta\omega \, .  
\end{equation}
 This result quantitatively explains the dependence of the optimal antibunching  (minimum in $g^{(2)}(0)$) 
 in figure~\ref{fig:gamma_pm}.
 At small pumping rate, the value of  $g^{(2)}(0)$ grows monotonously with $F$ \cite{savona10prl}. In this sense, the 
 smaller the pumping rate, the better the single-photon behavior of the output. However, in a practical device this
 would also lead to a vanishingly small number photons in the systems, and hence of single-photons emitted 
 from the cavity 1. In our conditions, the average number of photons $\langle \hat{a}^{\dagger}_1\hat{a}_1 \rangle$
 is always much smaller than 1, but on the order of $10^{-5} - 10^{-4}$, depending on the pumping strength,
 which means an expected single photon emission rate of $10-100$ kHz. 
 In terms of SPS efficiency, the possibility to enhance even further the single-photon emission rate (possibly beyond 
 the MHz) is an unequivocal advantage of the present source as compared to typical heralded 
 ones \cite{Fasel2004,kartik2012}.
 
A still open question, in view of actual applications of this scheme as a SPS, concerns the feasibility 
of pulsed operation, which would be required for single-photon emission on-demand. 
The delay-dependence of $g^{(2)}(\tau)$ is determined by beatings between the two oscillators. 
It shows pronounced oscillations at a rate determined by the tunnel coupling, 
and the condition $g^{(2)}(\tau)<1$ is fulfilled only over a delay on the order of 
$\tau\simeq \hbar/J$ \cite{savona10prl,bamba}, i.e. much shorter than the photon lifetime. 
This feature prevents the observation of photon antibunching in the pulsed excitation regime. 
However, it should be possible to address this issue by further processing the output of the two-cavity system. 
On one hand, a post-selective temporal filter with a $\hbar/J$ delay window can be applied at the output 
of the coupled cavity device to ensure the presence of single-photon Fock states. 
Moreover, it was already shown that photon antibunching can be optimized in similar situations 
through passive interferometry \cite{kimble92pra} or active quantum feedback schemes \cite{smith02prl}. 
In this sense, we believe $g^{(2)}(0) \ll g^{(2)}(\tau)$ is a necessary condition to be optimized before 
realizing a SPS, but further analysis is needed to characterize its output, depending on the specific 
device implementation. Such analysis goes beyond the scope of the present work.
}

\subsection{Effects of pure dephasing}

\begin{figure}[t]
\begin{center}
\includegraphics[width=0.5\textwidth]{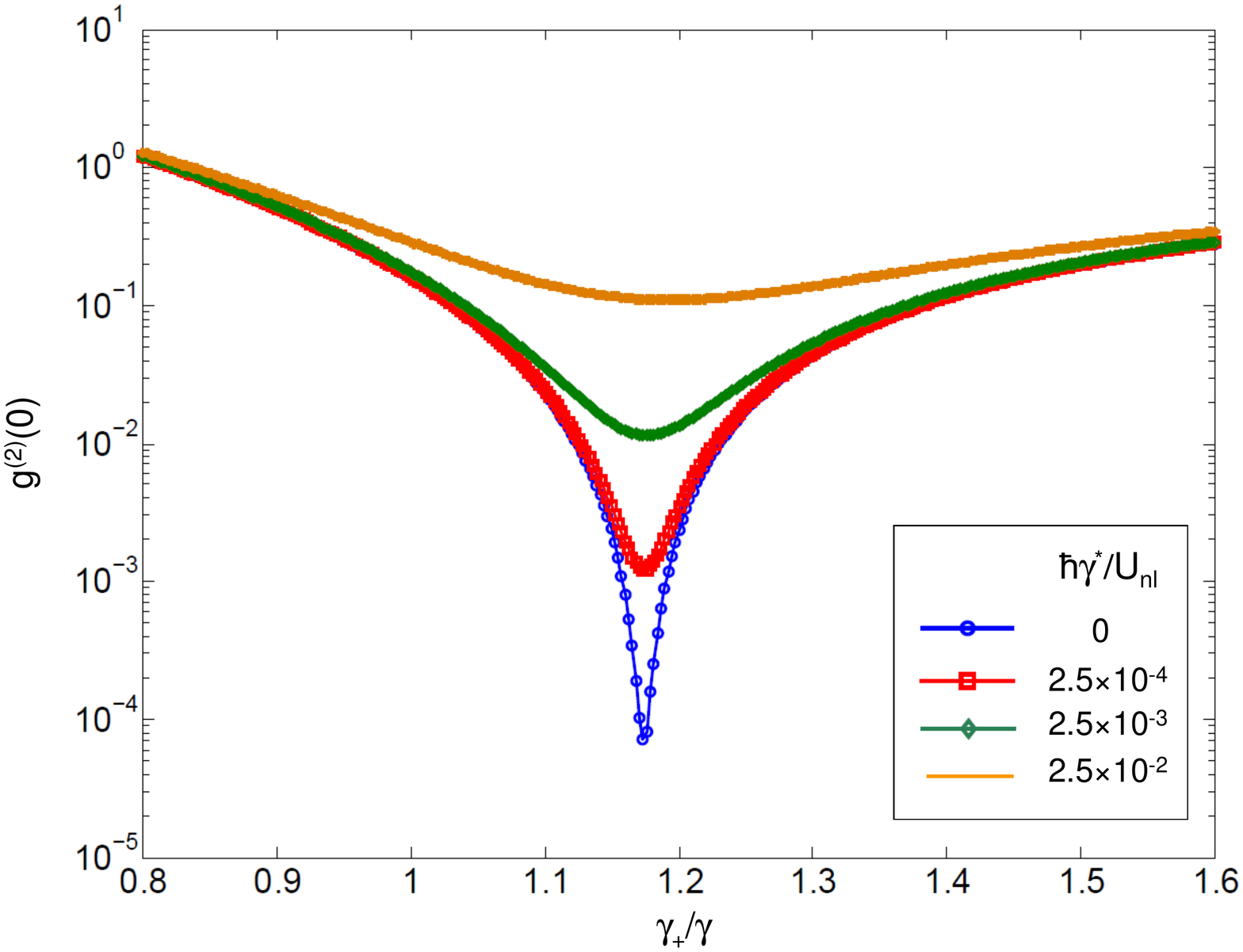}
\vspace{-0.5cm} 
\caption{Numerical solution for the zero-time-delay second-order correlation 
as a function of  $\gamma_{+}/\gamma$, for different values of pure dephasing 
rate (assumed equivalent for both symmetric and antisymmetric modes). Parameters 
are as in figure~\ref{fig:gamma_pm}, with $\gamma_{-}/\gamma=0.2$.} 
\label{fig:gamma_star}
\end{center}
\end{figure}

In photonic crystal molecules, the loss mechanism naturally leads to unbalanced 
quality factors for the symmetric and antisymmetric modes with respect
to the quality factor of the isolated cavity that can be on the order of 
$Q_- / Q_+ = \gamma_+ / \gamma_-  \sim 1 \to 10$.
Neglecting the counterintuitive effect of possible antisymmetric ground states \cite{caselli2012},
the lower energy mode is usually the symmetric combination of the two isolated cavity states, which 
has a reduced quality factor as compared to the isolated cavity \cite{atlasov08oe}.
Thus, we extract from the color plot in figure \ref{fig:gamma_pm} 
the curve corresponding to $\gamma_-= 0.2 \gamma$ (i.e. the quality factor of the 
antisymmetric mode is larger than the one of the isolated cavity).
The curve is plotted in figure \ref{fig:gamma_star} (curve corresponding to
$\gamma^{\ast}=0$), showing a close-to-ideal 
anti-bunching of $g^{(2)}(0) \simeq 7\times 10^{-5}$ for $\gamma_+ \simeq 1.2 \gamma$,
remarkably not far from the realistic ratio $\gamma_+ / \gamma_- \simeq 3$ 
reported in literature \cite{atlasov08oe} for coupled photonic crystal cavities.

A possible source of unwanted loss channels that could potentially spoil
the effects of photon blockade induced by tunnel-coupling is pure 
dephasing. In a semiconductor cavity, dephasing can be due to thermal 
fluctuations or other nonlinear mechanisms enhanced by the 
same electromagnetic field confinement, giving rise to index fluctuations 
within the cavity region and hence a pure dephasing rate. 
Even if such effect is usually neglected, and it is difficult to attribute a pure 
dephasing rate that is generally valid for any type of mechanism and material,
it is of utmost importance to estimate such detrimental effects on the SPS 
figure of merit for the purposes of the present work.   
As discussed in section \ref{sec:model}, the effects of pure dephasing can be simply 
modeled by solving the master equation after adding the Liouvillian term 
of the form represented in equation (\ref{eq:deph12}). 
We report in figure~\ref{fig:gamma_star} the calculated $g^{(2)}(0)$ 
as a function of $\gamma_{+}/\gamma$ for $\gamma_{-}/\gamma=0.2$ 
on increasing pure dephasing rate, $\gamma^{\ast}$. 
{ Suppression of antibunching occurs for values
of loss rates on the order of $\hbar\gamma^{\ast}\sim U_{\mathrm{nl}}$. }
Interestingly, the SPS figure of merit 
tolerance is such that at $\hbar\gamma^{\ast}/U_{\mathrm{nl}}\simeq 10^{-2}$ 
and  $\gamma_{+}/\gamma\simeq 1.2$, the system still displays an acceptable
single-photon nonlinear behavior  with $g^{(2)}(0) \simeq 0.1$, which confirms the 
robustness of such a scheme even with respect to loss channels that are difficult to 
control in an real experimental setting.
{ On the other hand, we expect that pure dephasing effects are most probably going to play a
marginal role in the present scheme, where small pump powers are required to efficiently
operate the SPS.}

\section{Conclusions}
\label{sec:conclu}

We have presented theoretical results supporting the proposal of a single-photon
source built on an integrated photonic platform made of passive material
components, where cavity-enhanced native nonlinearities are able to produce 
single-photon blockade. The building block is based on tunnel-coupled
resonators, or photonic molecules, coherently driven by an external laser source.
We have provided numerical evidence for the best combination of system parameters, 
such as cavity-laser detunings and modes losses, leading to optimal antibunching
of the output signal. 
In particular, we have found that a close to ideal single-photon output can be obtained 
from asymmetric losses of the coupled modes, as it naturally occurs in  
photonic crystal molecules.  
The robustness of such a device is tested against unwanted loss channels, e.g. giving 
rise to pure dephasing of the normal modes.

{ Further analysis will be needed to engineer the output of the device as a useful
single-photon wave packet for actual applications in integrated quantum circuits. }
However, the results presented here provide a useful starting ground to design highly innovative
quantum photonic devices based on integrated sources of pure single-photon states.
The main innovations come from the possibility to employ commonly nanostructured
materials from optoelectronic and CMOS-compatible industry, working at room temperature,
and in the telecom band for long distance low-loss data transmission. We believe this work
will further stimulate research in this direction, also in view of recent experimental demonstrations
of enhanced native optical nonlinearities in semiconductor-based photonic crystal platforms 
\cite{derossi2008,galli2010,graphene2012}.

\section*{Acknowledgements}
We acknowledge Timothy C. H. Liew for contributing to the early stages of  these calculations, and
for useful comments and suggestions, and L.C. Andreani, D. Bajoni, I. Carusotto, 
M. Galli, L. O'Faolain, T.F. Krauss, for stimulating discussions concerning the physics and the 
realization of this proposal. 
This work  was partly supported by NanoSci Era-NET project ``LECSIN'' coordinated by F. Priolo. 
One of the authors (S.F.) was supported by Fondazione Cariplo under project 2010-0523.




\begin{thebibliography}{99}
\bibitem{nota1}
We are assuming here that a tight-binding scheme for neighboring photonic resonators
holds, in which the tunnel coupling rate derives from the overlap between the evanescent 
tails of the cavity mode profiles. Such formal expression of the coupled modes is valid
only in the perfectly resonant condition, i.e. $\omega_1=\omega_2$, as we will assume
in the following of this work.
\bibitem{nota2}
Notice that we are neglecting self-consistent nonlinear effects on the 
cavity field profile induced by the Kerr-nonlinearity itself (e.g. field expulsion 
from the cavity region), which could renormalize the effective value of $U_{\mathrm{nl}}$.
\end{thebibliography}
\end{document}